\documentclass[twocolumn,showpacs,preprintnumbers,amsmath,amssymb,aps,prb]{revtex4}
\usepackage{graphicx}
\begin{document}

\title{
Dynamics and Separation of Circularly Moving Particles 
in Asymmetric Patterned Arrays    
} 
\author{
C. Reichhardt and C. J. Olson Reichhardt 
} 
\affiliation{
Theoretical Division,
Los Alamos National Laboratory, Los Alamos, New Mexico 87545 USA
} 

\date{\today}
\begin{abstract}
There are many examples of driven and active matter systems containing
particles that exhibit circular motion with different chiralities, such
as swimming bacteria near surfaces or certain types of self-driven colloidal 
particles. Circular motion of passive particles can
also be induced with an external rotating drive. 
Here we examine particles that move in circles and interact with a 
periodic array of asymmetric L-shaped 
obstacles. We find a series of dynamical phases as a function of swimming 
radius, including regimes where the particle motion is rectified, producing
a net dc motion.  The direction of the rectification varies with the swimming
radius, permitting the separation of particles with different swimming radii.
Particles with the same swimming radius but different chirality can also move
in different directions over the substrate and be separated.
The rectification occurs for specific windows of swimming radii corresponding
to periodic orbits in which the particles interact one or more times with
the barriers per rotation cycle.
The rectification effects are robust against the addition of thermal or
diffusive effects, and are in some cases even enhanced by these effects.
\end{abstract}
\maketitle

\vskip2pc

\section{Introduction}
Ratchet effects can occur for particles interacting with asymmetric substrates 
if the substrate is flashed on and off periodically or if an external ac
drive is applied, and result in
a net dc motion of the particles that is typically
aligned with the direction of the substrate asymmetry \cite{1}.
When the particles interact with each other as well as the substrate, the
ratchet effect can exhibit reversals where the particles move with the
substrate asymmetry in some regimes, and against it in others \cite{1}.
It is also possible to create ratchet effects for particles moving over 
two dimensional (2D) symmetric 
substrates by driving the particles with two
different ac drives that sum together to create 
an asymmetric pattern of particle motion
\cite{Reichhardt,2,R,3,4}. 
Directed motion on 
periodic substrates has been demonstrated experimentally 
for colloidal particles driven over magnetic substrates \cite{5,6}.

More recently, ratchet effects
have been observed in the absence of an external drive for systems of self-driven
particles, termed active matter.
Examples of active matter include swimming bacteria and 
crawling cells \cite{7,8,9}.  Numerous realizations
of artificial active matter have been created experimentally
for self driven colloidal systems \cite{10,11,12,13,14,15,16}.  
Active matter ratchet effects have been obtained for swimming bacteria
in an asymmetric funnel array in experiments \cite{8}, theory and simulations
\cite{17,19}, as well as for other types of swimming organisms\cite{18,21,22}
and eukaryotic cells \cite{9}. 
There has also been a recent proposal to rectify 
active particles interacting with symmetric substrates \cite{23}. 
Variants on the active matter ratchet have been used to create active 
matter-powered gears \cite{20}. 

The dynamics of active matter particles are often 
described by run and tumble 
motion where the particles
move for a persistence length or periodic time before reorienting, 
or the particles may move in a persistent random walk \cite{24,25}.      
In other active matter systems, the particles undergo 
circular motion such as found in  
swimming bacteria near surfaces \cite{31,32,33} and other 
types of swimming cells \cite{34,28}.
A variety of circle swimmers have been studied theoretically and in 
simulations \cite{26,27,30,29}, and in recent experiments, it was
demonstrated how to create       
artificial chiral colloidal moving particles 
that swim in circles with a fixed chirality
depending on the asymmetric nature of the particle itself \cite{35}. 
There are also proposals for creating molecular-sized chiral microswimmers 
by combining chiral molecules with chiral propellers \cite{36}. 
An understanding of the novel types of dynamics that arise for
circularly swimming particles interacting with patterned substrates could
lead to the ability to control the motion of such systems, which could be 
used for separation techniques or to extract work from active
matter.  Recently, it was shown that artificial swimmers in a 
periodic substrate array prefer to swim along certain 
symmetry directions of the array \cite{14}, while
Mijalkov and Volpe recently proposed a method for
sorting circle swimmers moving with opposite chirality by
using a substrate of chiral flower patterns or by having the particles 
move in channels of asymmetric arrays \cite{36}. 

In this work we examine circularly swimming particles 
interacting with an array of L-shaped barriers that create an asymmetric
landscape. 
The particles experience only contact interactions with the barriers, 
as in previous studies of run and tumble swimmers 
in asymmetric arrays \cite{17}.  After contacting a barrier, the particle
motion normal to the barrier is suppressed, and the particle swims along
the barrier at reduced speed until reaching the end of the barrier or
rotating away from it.
In the absence of a substrate, the
particles show no dc drift; however, when the array is present 
we observe a series of regimes in which the particles form periodic orbits
that produce net dc motion.
We find that the directed motion remains locked
over specific ranges of the swimming radii. In 2D,
the particle motion is generally locked along particular symmetry 
directions of the array, but the direction of the motion may switch as
the swimming radius varies.
We also find that for certain values of the swimming radius, 
there is no dc motion and the particle orbits are localized. 
The fact that the magnitude and the
direction of the dc motion can be controlled by varying the swimming radius
indicates that this type of substrate
could be used to separate particles with different swimming radii.
When the chirality of the swimmers is reversed, a different set of 
directed motion regimes appears, permitting
particles with the same swimming radius but different chirality 
to be separated.  
We find that the directed motion is generally reduced when the thermal
fluctuations increase; however, there are regimes where the directed
motion is enhanced by the additional fluctuations. 
We demonstrate these effects for oblique and square arrays 
of even L-shaped barriers as well as for 
a square array of simple one-dimensional (1D) barriers.   

\begin{figure}
\includegraphics[width=3.5in]{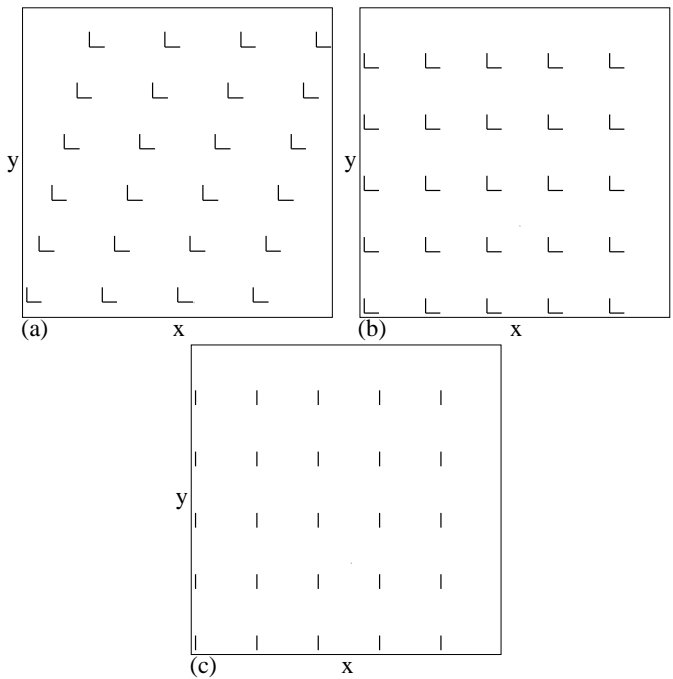}
\caption{ 
Sample geometries. (a) Oblique array of even L-shaped barriers. (b)
Square array of even L-shaped barriers.  (c) Square array of line barriers.
}
\label{fig:1}
\end{figure}

\section{Simulation and System}   
We consider a 2D system with periodic boundary conditions in the 
$x$ and $y$-directions
containing a periodic array of even L-shaped barriers, as illustrated in 
Fig.~\ref{fig:1}(a,b). 
In the even L shape, both arms of the L are identical in length.  We refer
to these barriers as ``L-shaped'' in the remainder of the paper.
The particles are initialized at random positions in the free regions
between barriers; in this work, we do not consider the effects of
particle-particle interactions.
The dynamics of particle $i$ is governed by the following 
overdamped equation of motion:  
\begin{equation}  
\eta \frac{d {\bf R}_{i}}{dt} = 
 {\bf F}_i^{m} + {\bf F}_i^{b} +  {\bf F}^{dc} +  {\bf F}^{T}_{i}. 
\end{equation} 
Here $\eta$ is the damping constant,  
${\bf F}^{m}$ is the motor force, ${\bf F}^{b}$ is the force from 
interactions with the barriers,
${\bf F}^{dc}$ is an external drift force, and 
${\bf F}^{T}$ is the thermal force.    
The motor force drives each particle in a circular manner
with ${\bf F}^{m}_i  = s_i[A_x\cos(\omega t){\hat {\bf x}} + 
A_y\sin(\omega t){\hat {\bf y}}]$, where $A\geq 0$ and $s_i=\pm 1$ is the
sign of the rotation.   
In the absence of a substrate, 
for $A_x=A_y=A$ and $s_i=1$ the particle moves in a counterclockwise 
orbit forming a circle with radius $A/\omega$; for $s_i=-1$, the motion is
clockwise. 
The barriers cause a short range steric repulsion of the particles,
and a particle in contact with a barrier will move along the barrier
with the component of ${\bf F}^m$ that is parallel to the barrier until
it reaches the end of the barrier or the orientation of ${\bf F}^m$ changes
enough to move the particle away from the barrier.
This is the same type of barrier interaction used in Ref.~\cite{17}, and
the sliding of particles along the barriers
is also consistent with experimental observations of swimming bacteria.
The thermal kicks come from the 
Langevin noise term ${\bf F}^{T}$ with the properties
$\langle F^{T}(t)\rangle = 0$ and 
$\langle F^{T}_{i}(t)F_j^T(t^{\prime})\rangle = 2\eta k_{B}T\delta_{ij}\delta(t - t^{\prime})$,   
where $k_{B}$ is the Boltzmann constant. 

We focus on systems of size $L \times L$ with
$L = 99$ in dimensionless units, and with barrier lattice constants of
around $a = 20$.  We have also run larger system sizes 
and find that all our results are robust.
We have studied different  lattice constants for fixed system size and  
find the same results; however, there is a systematic shift 
of the rectification regime to lower ac amplitudes as the barrier lattice 
constant is reduced. 
We consider three substrate types:
an oblique array of L-shaped barriers [Fig.~1(a)], 
a square array of L-shaped barriers [Fig.~1(b)],
and a square array of simple 1D barriers [Fig.~1(c)]. 
For the L-shaped barriers, 
the lack of rotational symmetry in the system produces
different motion for particles
with different chiralities, which is important for separation.
The system with simple 1D barriers has a rotational symmetry and is used
to clarify the effect of breaking rotational symmetry by comparison with
the L-shaped barrier systems.
We place $N = 980$ non-interacting particles in the sample
and 
measure the average particle velocities 
$\langle V_{x}\rangle = (1/N)\sum^{N}_{i = 1} {\bf v}_{i}\cdot {\hat {\bf x}} $ 
and $\langle V_{y}\rangle = (1/N)\sum^{N}_{i = 1}{\bf v}_{i}\cdot {\hat {\bf y}}$. 
At finite temperature, the use of this number of particles 
allows for sufficient averaging that 
a clear signature of a dc response can be obtained. 
If there is no 
net dc drift, then $\langle V_{y}\rangle = 0$ and $\langle V_{x}\rangle = 0$. 
In the absence of a substrate,
the average velocities are always zero.  

\begin{figure}
\includegraphics[width=3.5in]{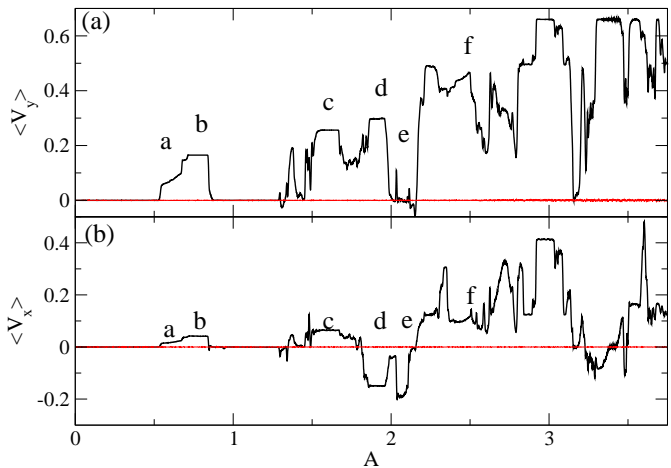}
\caption{ 
(a) $\langle V_{y}\rangle$ vs $A$ and 
(b) $\langle V_{x}\rangle$ vs $A$ 
for the system with the oblique L-shaped barrier array shown in 
Fig.~1(a) for counterclockwise swimming particles with $s_i=1$. 
The light (red) curves centered at $\langle V_{y}\rangle = 0$ and
$\langle V_{x}\rangle = 0$ 
are for the case where there is only an ac drive in the $y$-direction,
$A_y=A$ and $A_x=0$.
The dark (black) curves are for circular swimmers with $A_x=A_y=A$.
For the circular swimmers, there are a series of intervals of $A$ 
where there is a net dc motion, and in
several of these regions the velocities are constant or locked.       
The letters a-f indicate the values of $A$ where the particle trajectories 
in Fig.~4 were obtained.
}
\label{fig:2}
\end{figure}

\begin{figure}
\includegraphics[width=3.5in]{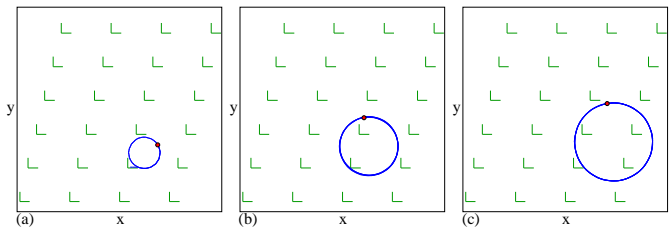}
\caption{
The trajectories of a single particle in the localized regimes for
the system in Fig.~\ref{fig:2} with counterclockwise swimming particles.
L-shapes: barriers; dot: individual particle; line: particle trajectory.
(a) $A=0.49$.  (b) $A=0.9$.  (c) $A=1.2$.
}
\label{fig:3}
\end{figure}

\begin{figure}
\includegraphics[width=3.5in]{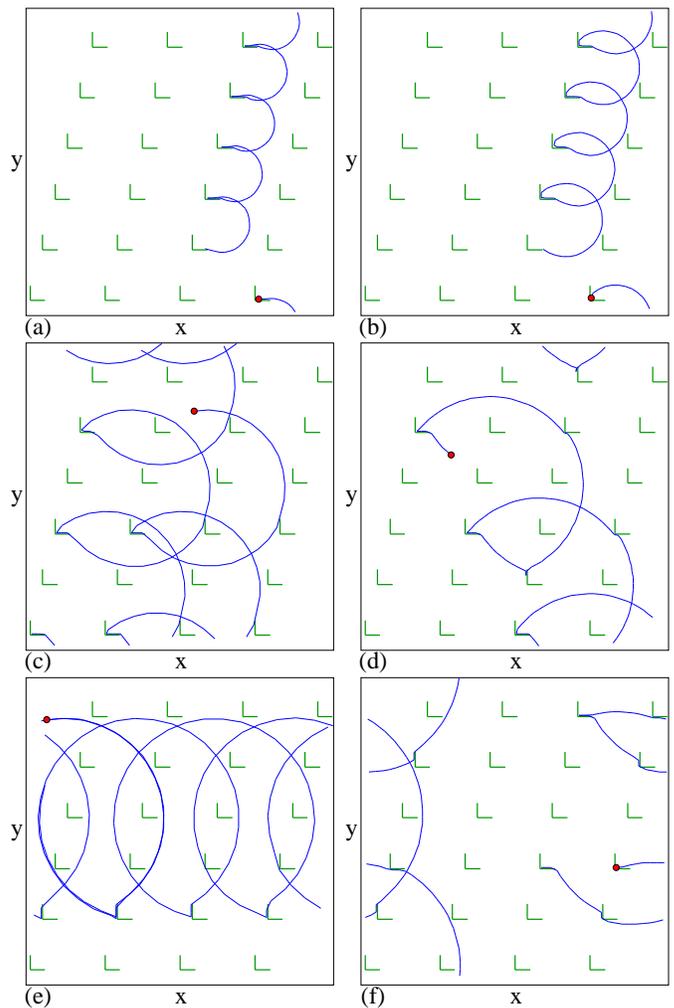}
\caption{
The trajectories of a single particle in selected rectification phases 
for the system in Fig.~2
with counterclockwise swimming particles.
The letters a-f in Fig.~2 correspond to the orbits shown in panels (a-f). 
L-shapes: barriers; dot: individual particle; line: particle trajectory.
At (a) $A = 0.57$, 
(b) $A=0.75$, and 
(c) $A=1.6$, the particle translates in the positive $y$ and $x$ directions 
and interacts with the barriers once per cycle.  
(d) At $A = 1.9$, the particle moves in the positive $y$ and 
negative $x$ directions and
interacts with the barriers twice per cycle. 
(e) At $A = 2.05$ the particle moves in the negative $x$ direction while
interacting with the barriers once per cycle. 
(f) At $A = 2.5$ the particle moves
in the positive $y$ and $x$ directions and interacts with the barriers 
three times per cycle. 
}
\label{fig:4}
\end{figure}

\section{Rectification Effects} 
We first focus on the oblique geometry shown in Fig.~1(a) with 
$s_i=1$ so that the
particles move in a counterclockwise direction.
In Fig.~2(a) we plot $\langle V_{y}\rangle$ vs $A$ for 
this system in the absence of thermal forces, $F^T=0$. 
As $A$ increases, the radius of the circular path that the particle would 
follow
in the absence of a substrate also increases. 
Fig.~2(b) shows the corresponding $\langle V_{x}\rangle$ vs $A$. 
We also
plot $\langle V_{x}\rangle$ and $\langle V_{y}\rangle$ vs $A$ 
for particles with $A_x=0$ and $A_y=A$ that only move up and down but do not
rotate.
For the 1D driving, there is no directed motion, as indicated by  
the curves centered at $\langle V_{x}\rangle = 0.0$ and 
$\langle V_{y}\rangle = 0.0$ in 
Fig.~2(a,b).  
For the circular swimmers, there are numerous intervals of $A$ where there
is a net dc flow of particles. 
For $ 0 < A < 0.52$, there is no dc response since the particle 
orbits are small enough that the particles can move between
the barriers without contacting them, as illustrated in
Fig.~3(a) at $A = 0.49$. 
For $0.52 < A < 0.86$, Fig.~2(a,b) indicates that
there is net dc motion in both the positive $y$ 
and positive $x$ directions, with a larger dc response
in the $y$ direction, $\langle V_y\rangle > \langle V_x\rangle.$
This interval also includes
a plateau region in $\langle V_{y}\rangle$ over which the velocity
remains nearly constant,
similar to the mode locking 
phenomenon found for particles driven with asymmetric drives over
symmetric periodic substrates \cite{2}.
In this regime, as illustrated 
in Fig.~4(a) 
at $A = 0.57$, 
individual particles fall into a periodic orbit that contacts the
barrier once per drive cycle.  When the particle is in contact with the barrier,
it moves along the bottom of the L shape and is temporarily trapped in
the corner of the L until its motion reverses and it moves up by one
barrier in the $y$ direction.
During successive orbits, the particle channels along the symmetry 
direction of the lattice so that it moves over by
one column in the positive $x$-direction after six drive cycles. 
At $A=0.75$, Fig.~4(b) shows 
that the particles are guided along the barriers near the corner of the L.
We observe the ratio $\langle V_y\rangle/\langle V_x\rangle=4.0$ at
$A=0.75$ since
the particles move six lattice constants in the $y$ direction for every
1.5 lattice constants in the $x$ direction in this locked motion regime.

At $A = 0.9$, $\langle V_{y}\rangle$ and $\langle V_{x}\rangle$ are zero 
again in Fig.~2, and in Fig.~3(b) we show the corresponding localized particle
orbit in which the particle encircles two barriers without touching
them.
At $A = 1.2$ there is still no net dc signal, and the localized 
particle orbit now encircles three barriers in a single cycle 
as shown in Fig.~3(c). For $ 1.3 < A < 1.8$,
a dc rectification occurs 
where  the particles translate in both the 
positive $x$ and $y$ directions. 
For the velocity plateau region centered
at $A = 1.6$, $\langle V_{y}\rangle/\langle V_{x}\rangle = 5.0$, 
and Fig.~4(c) shows that in this region
the particle interacts with 
a single barrier per cycle and also moves past multiple barriers on each cycle. 
For $1.8 < A < 2.0$ we observe
a rectification region  
where the particles move in the positive $y$ and the
negative $x$ directions
with $\langle V_{y}\rangle/|\langle V_{x}\rangle| = 2.0$.  In
this region, as illustrated in 
Fig.~4(d) 
for $A = 1.9$, the 
particles interacts with three barriers per cycle. 
The first interaction is with the front of one L-shaped barrier, the
second is with the back of another barrier, and the third is with the
back corner of yet another barrier.

There are also intervals of $A$ where the particles move strictly in the 
negative $x$-direction, such as for 
$ 2.035 < A < 2.1$. Figure~4(e) shows 
that in this regime at $A = 2.05$,
the particle interacts with the back of one L-shaped barrier 
per cycle and translates in the negative $x$ direction. 
For higher values of $A$, more complex motions occur.  For example, at 
$A = 2.25$, illustrated in Fig.~4(f), the particles move in the positive
$x$ and $y$ directions with
$\langle V_{y}\rangle/\langle V_{x}\rangle = 4.0$, and
interacting with four barriers per cycle.
For increasing $A$, additional rectifying orbits appear that
become increasingly complex as the particle can interact with larger 
numbers of barriers per cycle. For most applications,
the first few rectifying orbits will likely be the best regimes to consider.  

In addition to integer ratios of $\langle V_y\rangle/\langle V_x\rangle$,
we also observe simple noninteger ratios such as 
$\langle V_y\rangle/\langle V_x\rangle=1.5$.
In general, for the oblique lattice geometry and 
counterclockwise swimming particles, $\langle V_y\rangle$ is
larger than $\langle V_x\rangle$.
We have only observed very small windows over which the particles
move in the negative $y$ direction, and we found no regimes where
the particle motion is strictly along the $y$ direction, due to the
fact that the easy channel direction along the symmetry axis of the
oblique lattice is not aligned with the $y$ direction.
The fact that the particles move in different directions 
for different values of $A$ or orbit radii indicates that
particles with different swimming radii but the
same chirality could be sorted using the oblique substrate.
 
There are several differences between this system and the dc rectification
induced by driving ac particles over symmetric periodic substrates
\cite{Reichhardt,2}. 
In the previous work, the substrates were modeled as smooth egg-carton type
potentials induced by pinned particles that repelled the remaining
particles via a Yukawa interaction potential.
In that egg-carton system, the mobile particles are always interacting with 
the potential substrate, as appropriate for modeling colloidal particles on
periodic optical trap arrays \cite{37}.  
In the present work, the substrate is flat except for the L-shaped barriers,
and the particles experience only contact interactions with the barriers.
In the egg-carton system under a circular ac drive, a transverse dc motion 
occurs only if an
additional longitudinal dc drift is applied to the particles \cite{Reichhardt};
if a noncircular ac drive is applied that produces a particle orbit with
broken spatial symmetry, then dc rectification can
occur even without a dc drift drive \cite{2,3,4}.
In the present work, the ac driving orbits are symmetric, and the barrier 
lattice
introduces the asymmetry required for rectification to occur.
The system we propose here should be very straightforward to implement due
to the simple barrier shapes and particle interactions.

\subsection{Opposite Chirality}

\begin{figure}
\includegraphics[width=3.5in]{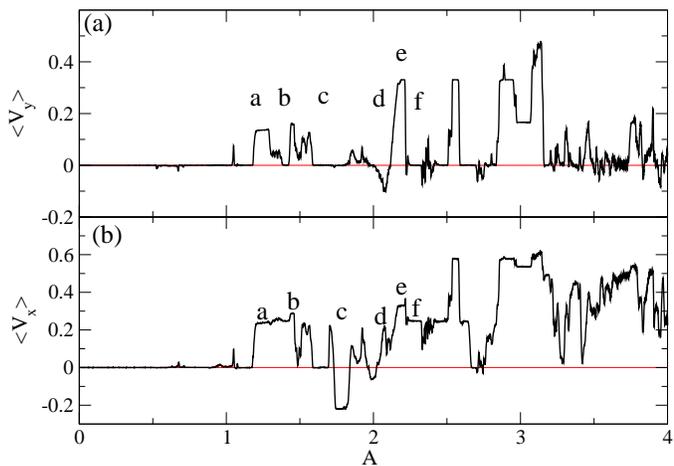}
\caption{
(a) $\langle V_{y}\rangle$ vs $A$
and (b) $\langle V_{x}\rangle$ vs $A$ for the system
from Fig.~1(a) with an oblique lattice of L-shaped barriers 
for clockwise swimming
particles with $s_i=-1$.  
The $\langle V_y\rangle=0$ and $\langle V_x\rangle=0$ lines are
drawn as a guide to the eye.
Here, we generally find $\langle V_x\rangle>\langle V_y\rangle$, the reverse
of what is shown in Fig.~2 for counterclockwise swimming particles.
The letters a-f indicate the values of $A$ where the particle trajectories
in Fig.~6 were obtained.
}
\label{fig:5}
\end{figure}

\begin{figure}
\includegraphics[width=3.5in]{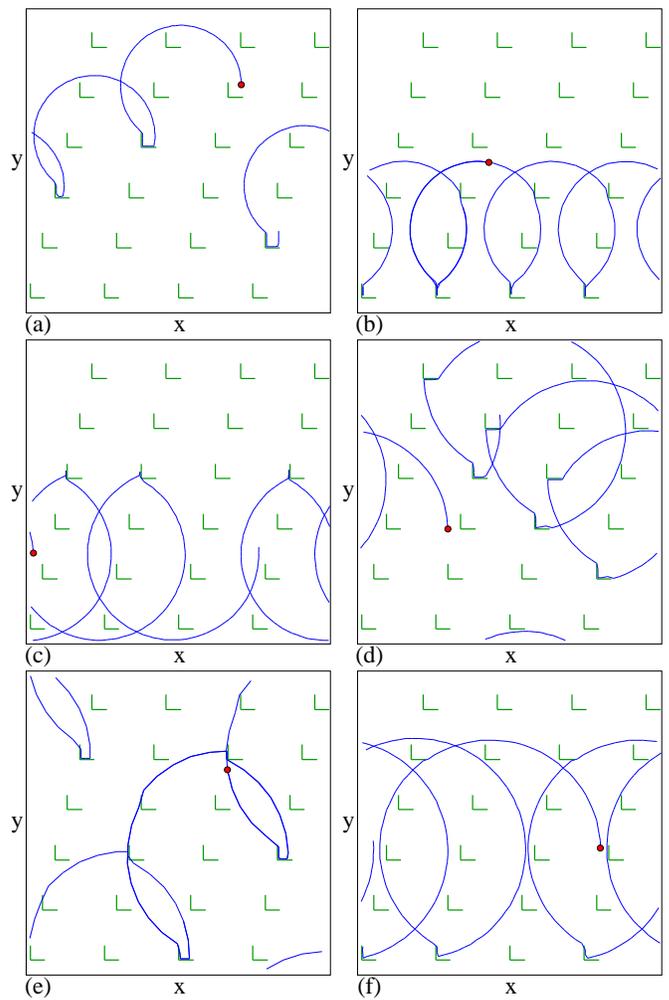}
\caption
{The trajectories of a single particle in selected rectification phases for
the system in Fig.~5 for clockwise swimming particles with
$s_i=-1$.
The letters a-f in Fig.~5 correspond to the orbits shown in panels (a-f).
L-shapes: barriers; dot: individual particle; line: particle trajectory.
(a) At $A = 1.25$, the particle translates in the positive 
$x$ and $y$ directions. 
(b) At $A = 1.4$, the particle translates only in the positive $x$ direction.
(c) At $A = 1.77$, the particle translates only in the negative $x$ direction.
(d) At $A = 2.08$, the particle translates in the positive $x$ direction and 
negative $y$ direction.
(e) At $A = 2.15$, the particle translates in the positive $x$ and $y$
directions.
(f) At $A = 2.28$, the particle translates only in the positive $x$-direction.
}
\label{fig:6}
\end{figure}

We next consider the same oblique array of L-shaped barriers 
but reverse the sign of the drive by setting $s_i=-1$ 
so that the particles move clockwise.
In Fig.~5(a,b) we plot $\langle V_{y}\rangle$ and $\langle V_{x}\rangle$ 
versus $A$, where we generally find $\langle V_y\rangle < \langle V_x\rangle$,
the opposite of what occurred for the counterclockwise swimming particles.
There is also no dc rectification for $A<1.2$ for 
$s_i=-1$, whereas for the counterclockwise rotation rectification occurred
down to $A=0.52$.
Figure 5 shows that for
$1.2 < A < 1.588$, the particles rectify in the positive $x$ and $y$
directions.
The trajectory of a particle in this regime is shown in Fig.~6(a) for 
$A=1.25$,
where the particle interacts with  almost the entire length of the front side 
of the barrier, producing a net translation in the 
positive $x$ and positive $y$ directions. 
At $A = 1.4$, $\langle V_{y}\rangle = 0$ and the velocity in the 
$x$-direction is positive.  Here,
Fig.~6(b) shows that the particle interacts with two barriers per cycle,
once with the front part of the barrier and once with the back part of the
barrier.
There is an interval of $A$ centered around $A = 1.77$  where the 
particles move only in the negative $x$-direction, and
Fig.~6(c) indicates that here the particle encounters only one barrier per
cycle and interacts with the back side of the L shape.
Near $A = 2.08$ there is a regime where the particle moves in the
positive $x$ direction and the negative $y$ direction, as shown by the
particle orbits in Fig.~6(d).
Around $A = 2.15$ the particle moves in both the positive 
$x$ and $y$ directions again, as highlighted 
in Fig.~6(e) where the particle interacts with three barriers per cycle. 
In the region near $A = 2.28$, Fig.~6(f) shows that
the particle translates only in the 
positive $x$-direction and interacts
with just one barrier per cycle in a manner similar to that found
for $A = 1.4$ [Fig.~6(b)], but in this case the orbits form a much wider
arc.
For higher values of $A$ similar sets of rectifying motions occur,
with the particles typically translating a further distance per cycle, 
as indicated by
the increase in $|\langle V_{x}\rangle |$ and $|\langle V_{y}\rangle |$ 
at higher $A$ in Fig.~5.   

\begin{figure}
\includegraphics[width=3.5in]{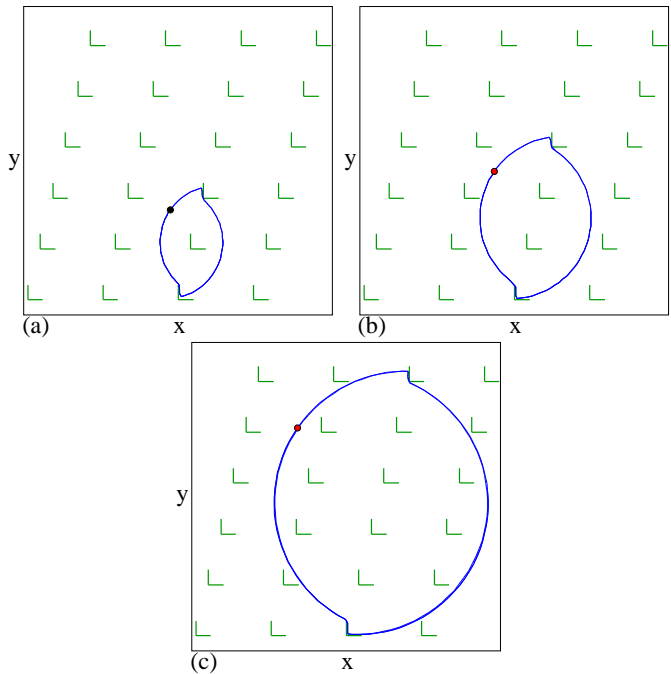}
\caption{
The trajectories of a single particle in the localized regimes for the
system in Fig.~5 with clockwise swimming particles.
L-shapes: barriers; dot: individual particle; line: particle trajectory.
(a) $A = 1.15$. (b) $A = 1.65$. (c) $A = 2.68$.
}
\label{fig:7}
\end{figure}

In the intervals of $A$ where there is no net motion in Fig.~5, the 
particles adopt localized orbits; however, unlike the
counterclockwise case shown in Fig.~3 where the particles undergo circular 
orbits that do not interact with the barriers, 
for the clockwise orbits the particles can interact with the barriers but
still produce no net translation.
For $A < 0.5$ the particle orbits are small enough to fit between the
barriers without contacting them, while for $A \geq 0.5$ the orbits
can touch the barriers,
as shown in Fig.~7(a) for $A = 1.15$ where the particle interacts with 
the back of one L and then hits the front of another L while encircling a
third barrier. 
At $A = 1.65$ the motion is localized and Fig.~7(b) shows that each
particle forms
an elliptical orbit similar to the one at $A = 1.15$ where the
particle interacts with two barriers during each cycle; in this case, however,
the orbit encircles two barriers.
For $A = 2.68$, illustrated in Fig.~7(c), the same type of orbit occurs 
but the particle encircles ten barriers per cycle.  

\begin{figure}
\includegraphics[width=3.5in]{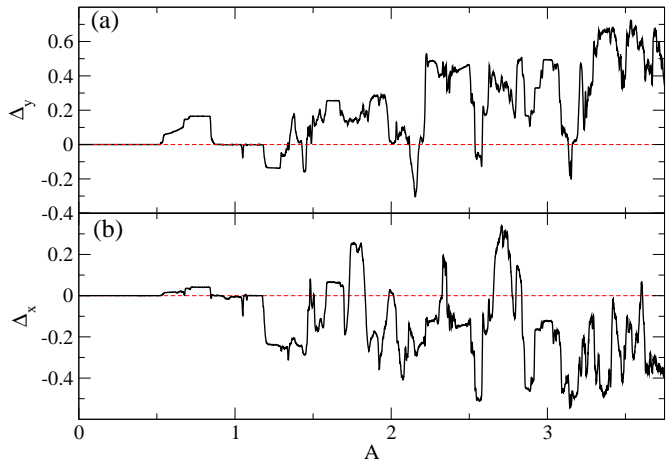}
\caption{
(a) $\Delta_y$, the difference between $\langle V_y\rangle$ for the 
counterclockwise and clockwise rotating particles, vs $A$ in samples
with an oblique lattice of L-shaped barriers.
(a) $\Delta_x$ vs $A$.
In the regions with non-zero mean values of $\Delta_x$ or $\Delta_y$,
particles with different swimming chiralities move at different speeds
and/or in different directions.     
}
\label{fig:8}
\end{figure}

Since the different chiralities produce different modes of motion, the
oblique lattice of L-shaped barriers
can be used to separate particles moving with different chiralities. 
In Fig.~8 we plot the velocity differences
$\Delta_y=\langle V_y\rangle_{s_i=1} - \langle V_y\rangle_{s_i=-1}$
and $\Delta_x=\langle V_x\rangle_{s_i=1} - \langle V_x\rangle_{s_i=-1}$, where
the velocity of the clockwise swimmers is subtracted from the velocity of
the counterclockwise swimmers.
Whenever $\Delta_y$ and/or $\Delta_x$ have a finite value,
particles with different chiralities move at different speeds and/or
in different directions, and separation of the particles can be achieved.

\begin{figure}
\includegraphics[width=3.5in]{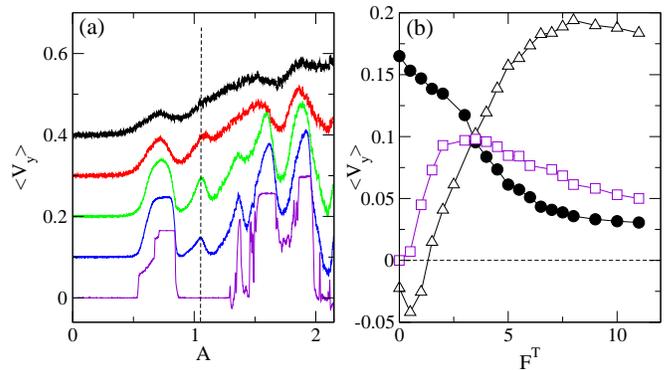}
\caption{
(a) $\langle V_{y}\rangle$ vs $A$ for the system in Fig.~2 with
an oblique lattice of L-shaped barriers and
counterclockwise moving particles ($s_i=1$) at different thermal force values 
$F^{T} = 0.0$, 1.0, 2.0, 4.0, and $6.0$, from bottom to top.  The
curves have been successively shifted up for clarity. 
The dashed line highlights the appearance of a new peak
near $A = 1.05$ for finite temperature.
(b) $\langle V_{y}\rangle$ vs $F^{T}$ for 
$A = 0.75$ (circles), which monotonically decreases;
$A = 1.05$ (squares), which starts at $\langle V_{y}\rangle = 0$, 
reaches a maximum and then decreases; and
$A = 2.12$ (triangles), which starts with  $\langle V_{y}\rangle <0$, 
reverses to a positive value, and then
decreases at higher $F^{T}$.  
}
\label{fig:9}
\end{figure}

\begin{figure}
\includegraphics[width=3.5in]{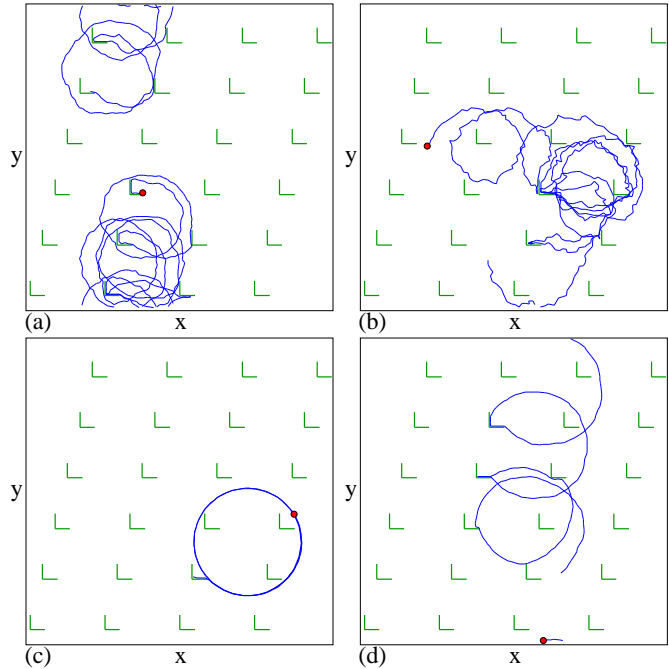}
\caption{
The trajectories of a single particle for the system
in Fig.~9 with counterclockwise swimming particles and finite temperature.
L-shapes: barriers; dot: individual particle; line: particle trajectory.
(a) $A = 0.9$ and $F^{T} = 4.0$. (b) $A = 0.9$ and $F^{T} = 8.0$.
(c) $A = 1.05$ and $F^{T} = 0$. (d) $A = 1.05$ and $F^{T} = 2.0$. 
}
\label{fig:10}
\end{figure}

\section{Thermal Effects}

In Fig.~9(a) we plot $\langle V_{y}\rangle$ versus $A$ for the system
from Fig.~2 with an oblique lattice of L-shaped barriers and
counterclockwise swimmers 
for different thermal forces $F^T=0$ to 6.0. 
In general we find that over ranges of $A$ where plateaus occur at
$F^T=0$, $\langle V_y\rangle$ decreases in magnitude and the plateau
features smear out as $F^T$ increases.
For ranges of $A$ in which $\langle V_y\rangle=0$ in the absence of
thermal fluctuations, the value of $\langle V_y\rangle$ becomes finite
when $F^T$ is raised above zero.
Here, the localized particle orbits become smeared by thermal fluctuations
and the particles occasionally interact with extra barriers, producing a
nonzero amount of rectification.
The magnitude of the peaks in $\langle V_y\rangle$ is suppressed by
temperature 
since the thermal kicks can 
temporarily cause the particles to deviate from their
rectifying orbits. 
For increasing $F^T$, 
we find some rectification for most values of $A$, as shown in
Fig.~9(a).
For example, at $A = 0.9$ there is no rectification in the
absence of thermal fluctuations; however, 
for $F^{T} = 4.0$ there is some rectification. 
Thermally induced rectification is illustrated in Fig.~10(a) for
$A=0.9$ and $F^T=4.0$, where the particle is localized during some time
intervals and undergoing net transport in the positive $y$ direction during
other time intervals.
For high $F^T$, we observe a general decrease in the magnitude of the
rectification when the particle trajectories become increasingly random,
as shown in
Fig.~10(b) for
$A = 0.9$ and $F^{T} = 8.0$. 
Over certain ranges of $A$, the addition of thermal effects create
new locking effects that were not present in the zero temperature system.
An example of this is highlighted in 
Fig.~9(a) with a dashed line near $A = 1.05$. 
Here, at $F^{T} = 0$, $\langle V_y\rangle=0$, 
but at finite temperature a peak in $\langle V_{y}\rangle$ emerges. 
This peak persists as the temperature increases before 
gradually smearing out at high temperatures. 
In this case,
at  $F^{T} = 0$ and $A = 1.05$, Fig.~10(c) indicates that the particle moves
in a circle that just misses interacting with the barriers during each 
cycle, 
while when $F^{T} = 2.0$  
for the same value of $A$, Fig.~10(d) shows that the particles now have a 
much higher probability of interacting with the barriers. 
As a result, a finite translation in the positive $y$ direction occurs
only for finite temperature.
 
In Fig.~9(b) we plot $\langle V_{y}\rangle$ versus $F^{T}$ 
at specific values of $A$ for the system 
in Fig.~9(a) to highlight the different thermal behaviors.  
At $A=0.75$, which is
on the first finite rectification plateau, 
$\langle V_{y}\rangle$ decreases monotonically with increasing 
$F^T$.
At $A = 1.05$, initially $\langle V_{y}\rangle = 0,$ and 
$\langle V_y\rangle$ gradually increases with increasing $F^T$ 
to a maximum value before
decreasing at higher thermal force. 
At $A = 2.12$, $\langle V_y\rangle$ is negative for $F^{T} = 0,$
and as the thermal force increases, there is a rectification reversal from 
negative to positive values of $\langle V_y\rangle$, 
which reaches a maximum and then slowly decreases for larger thermal force.     

\begin{figure}
\includegraphics[width=3.5in]{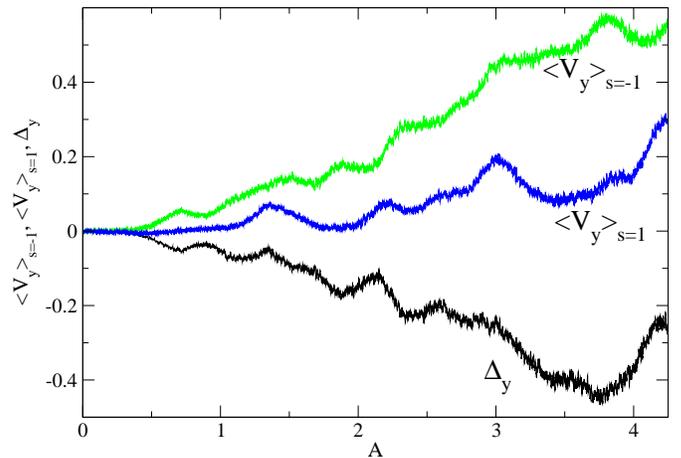}
\caption{
Samples with an oblique lattice of L-shaped barriers at $F^T=6.0$.
Upper curve: $\langle V_{y}\rangle_{s_i=-1}$ (clockwise particles) vs $A$. 
Center curve: $\langle V_{y}\rangle_{s_i=1}$ (counterclockwise particles) vs $A$. 
Lower curve: The corresponding $\Delta_y$ vs $A$.
}
\label{fig:11}
\end{figure}

In Fig.~11 we plot 
$\langle V_{y}\rangle_{s=-1}$ and
$\langle V_{y}\rangle_{s=1}$ 
vs $A$ for the clockwise and counterclockwise swimming particles,
respectively,
as well as the difference $\Delta V_y$, 
which indicates that even at higher temperatures, 
particles of different chiralities can move in 
different directions so that separation is still possible.    

\begin{figure}
\includegraphics[width=3.5in]{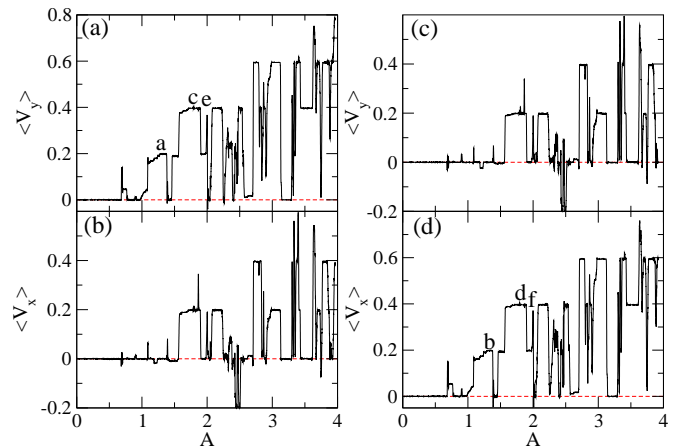}
\caption{
(a) $\langle V_y\rangle$ vs $A$ and
(b) $\langle V_x\rangle$ vs $A$ for the system with the square
L-shaped barrier array shown in Fig.~1(b) for counterclockwise swimming
particles with $s_i=1$.  
(c) $\langle V_y\rangle$ vs $A$ and
(d) $\langle V_x\rangle$ vs $A$ for the same 
system for clockwise swimming
particles with $s_i=-1$.  
The $\langle V_y\rangle=0$ and $\langle V_x\rangle=0$ lines are drawn as
a guide to the eye.
The curves for (a) and (d) are indistinguishable from each other,
as are the curves in (b) and (c), 
indicating that the velocity response is interchanged when the chirality
is reversed.
The letters a-f correspond to the 
values of $A$ at which the orbits in Fig.~13 were obtained.
}
\label{fig:12}
\end{figure}

\begin{figure}
\includegraphics[width=3.5in]{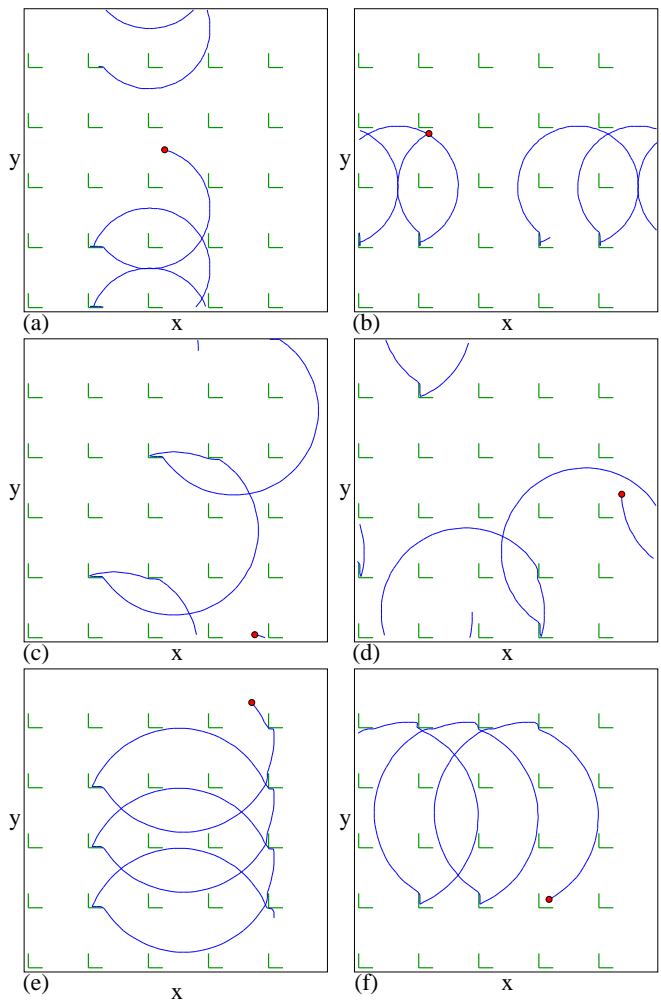}
\caption{
The trajectories of a single particle in selected rectification phases
for the system in Fig.~12 with 
a square lattice of L-shaped barriers.
(a) Counterclockwise swimmers at $A=1.25$ translate in the $y$ direction.
(b) Clockwise swimmers at $A=1.25$ translate in the $x$ direction.
(c) Counterclockwise swimmers at $A =  1.75$. 
(d) Clockwise swimmers at $A = 1.75$.
(e) Counterclockwise swimmers at $A  = 1.95$ translate in the $y$ direction.
(f) Clockwise swimmers at $A = 1.95$ translate in the $x$ direction.   
}
\label{fig:13}
\end{figure}

\section{Other Geometries}
We next consider the case of square lattices of L-shaped barriers, 
as illustrated in Fig.~1(b). 
In Fig.~12(a,b) we plot $\langle V_{y}\rangle$ and $\langle V_{x}\rangle$ 
versus $A$ for this system with counterclockwise moving particles. 
We find a set of rectification regimes as a function of $A$ that
are similar to those in the system with the oblique array of L-shaped
barriers, although one notable difference 
is that for the square lattice it is possible to have 
dc motion oriented strictly in 
the $y$ direction, such as at $A = 1.35$. 
As was the case for the oblique array, the square array produces
some intervals of $A$ where  the
motion is in the negative $x$-direction, 
such as near $A = 2.5$. 
When the chirality of the swimmers is reversed in the square array,
Fig.~12(c) shows that 
$\langle V_y\rangle$ versus $A$ for the clockwise swimmers
has the exact same form as 
$\langle V_x\rangle$ versus $A$ shown in Fig.~12(b) for the counterclockwise
swimmers.
Similarly, 
$\langle V_x\rangle$ versus $A$ in Fig.~12(d) for the clockwise swimmers
has the same form as 
$\langle V_y\rangle$ versus $A$ in Fig.~12(a) for the counterclockwise
swimmers.
This symmetry in the velocity response is a result of the 
higher symmetry of the square lattice, which causes the combination of
a reversal of the chirality of the swimmer and a 90-degree rotation of the
substrate to appear the same as the unrotated 
and unreversed system, swapping the
$x$ and $y$ velocity responses.  A similar symmetry is not present in the
oblique lattice.
The exchange of the velocity response from 
$\langle V_y\rangle$ to $\langle V_x\rangle$ with the change in chirality
makes the square barrier lattice more convenient than the oblique lattice
for particle separation techniques since a measurement of the motion of one
chirality of swimmer immediately determines how a swimmer of the opposite
chirality will move.
For example,
if one chirality of swimmer moves strictly in the
$y$ direction at a specific value of $A$, 
then the opposite chirality of swimmers must move 
strictly in the $x$ direction for the same value of $A$. 
This is illustrated for $A = 1.25$ in Fig.~13(a), which shows the
counterclockwise swimmers translating in the $y$ direction, and Fig.~13(b),
which shows the clockwise swimmers undergoing the same motion rotated
by 90 degrees, with a resulting translation in the $x$ direction.
Similarly, at $A = 1.95$, Fig.~13(e) shows that the 
counterclockwise swimmers translate in the $y$ direction while,
in Fig.~13(f), the clockwise swimmers translate in the $x$ direction 
with an orbit of the same shape but rotated by 90 degrees.
At $A = 1.75$, Fig.~12(a,b) shows that counterclockwise swimmers
have positive velocities in
both the $x$ and $y$ directions with $\langle V_y\rangle > \langle V_x\rangle$.
In the corresponding particle trajectory in Fig.~13(c), 
the particle moves further in $y$ than in $x$ after several cycles. 
For the clockwise swimmers at $A=1.75$,
the motion is similar with positive velocities in both the $x$ and $y$
directions, but now $\langle V_y\rangle<\langle V_x\rangle$, and the
corresponding trajectory in Fig.~13(d) is a rotated version of the
trajectory in Fig.~13(c).

For both the square and oblique lattices of L-shaped barriers,
we also find that for intervals of $A$ where there
is no rectification, 
the application of a dc drift force in one of the directions can induce
locking of the motion in the driven direction as well 
as a transverse rectification. There are also 
smaller regimes of the external drive 
where the transverse rectification occurs at different values of the
external drive for particles of different chirality, indicating that this
could provide another method for particle separation.

\begin{figure}
\includegraphics[width=3.5in]{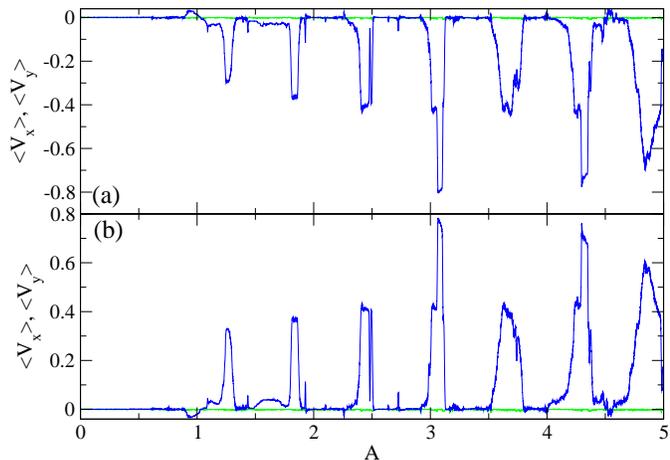}
\caption{
Sample with a square array of line barriers illustrated in Fig.~1(c).
(a) $\langle V_{x}\rangle$ (lower dark curve) and 
$\langle V_{y}\rangle$ (upper light curve) vs $A$ 
for counterclockwise swimmers.
(b) $\langle V_{x}\rangle$ (upper dark curve) and 
$\langle V_{y}\rangle$ (lower light curve) vs $A$ 
for clockwise swimmers.
For both types of swimmers, $\langle V_y\rangle \approx 0$ over the
entire range of $A$.
}
\label{fig:14}
\end{figure}

\begin{figure}
\includegraphics[width=3.5in]{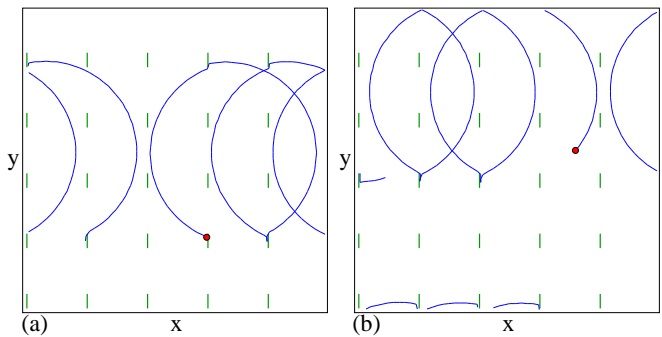}
\caption{
The trajectories of a single particle in selected rectification phases
for the system in Fig.~15 with a square lattice of 1D barriers at
$A=1.85$.
(a) Counterclockwise swimmers translate in the negative $x$ direction.
(b) Clockwise swimmers translate in the positive $x$ direction.  
}
\label{fig:15}
\end{figure}

We have also considered particles moving in a square array of 1D 
barriers as illustrated in Fig.~1(c). 
This system has an even higher degree of symmetry than the square lattice
of L-shaped barriers, since the barrier lattice now has a rotational symmetry.
In Fig.~14(a) and (b) we plot $\langle V_{x}\rangle$ 
and $\langle V_{y}\rangle$ vs $A$ for the array of 1D barriers for
counterclockwise and clockwise swimmers, respectively.
In both cases we observe rectification in the $x$-direction 
while $\langle V_{y}\rangle \approx 0.$ 
The rectification regimes appear at regularly spaced intervals
in $A$, and the rectification is in the negative $x$ direction for the
counterclockwise swimmers and in the positive $x$ direction for the
clockwise swimmers.
Figures 15(a) and (b) illustrate this rectification for the counterclockwise
and clockwise swimmers, respectively, at $A=1.85$
where the particles interact with
two barriers per cycle. 
Both chiralities of swimmers execute orbits with the same shapes, but
reversing the chirality flips the orbit by 180 degrees.
Although the regime where rectification occurs for the 1D barriers
is smaller than that for the L-shaped barriers, the
fact that particles of different chirality move in opposite directions through
the 1D barriers could make the 1D barrier array the most
practical choice for a separation technique.

\section{Summary}
We have examined circularly swimming particles 
interacting with 
periodic arrays of L-shaped or rod-shaped barriers. 
The particles experience only
contact interactions with the barriers, and when in contact with the
barrier, a particle moves according to its driving force component that
is parallel to the barrier wall.
In the absence of barriers, there is no net dc motion.
For oblique and square arrays of L-shaped barriers,
we observe a rich variety of distinct rectifying phases where dc motion
can arise in numerous different directions.
These dynamic phases appear 
as the swimming radius of the particles is varied, and
the rectifying phases are associated with
periodic orbits in which 
the particles interact with one
or move barriers during each swimming cycle. 
At certain radii values, 
the particles 
become localized and exhibit no net dc rectification.    
Since the rectification direction varies with swimming radius,
such barrier arrays could be used to separate
particles with different swimming radii but the same chirality of
swimming direction. 
When the chirality is reversed, the particles rectify in a different 
set of directions as the swimming radius varies,
implying that particles with different chirality can 
also be separated with the barrier arrays. 
Our results are robust at finite temperature, and we find that
the thermal fluctuations can in some cases enhance the rectification 
by increasing the frequency with which the particles
interact with the barriers.   
For square arrays of L-shaped barriers,
the velocity response of counterclockwise swimming particles is rotated
with respect to that of clockwise swimming particles, with the $y$ ($x$)
response of the counterclockwise swimmers becoming the $x$ ($y$) response of
the clockwise swimmers.
This indicates that in a regime
where the counterclockwise swimmers translate strictly in the 
positive $x$ direction, the clockwise
swimmers move strictly in the positive $y$ direction. 
The symmetry of the velocity response could be used to more
readily separate particles of different chiralities. 
For a square array of 
1D barriers, 
we find that rectification effects
still occur over reduced ranges of the swimming radius.
For the 1D barrier arrays, the rectification occurs 
only in one direction for a given chirality, and particles of the opposite
chirality move in the opposite direction.
Our results could be realized experimentally using 
the recently studied artificial circle swimmers, swimming
bacteria, or other active matter particles that move in circles. 
Beyond self-driven particles, it should also be possible to 
create systems of particles that move in circles due to some type
of rotating external field.

\acknowledgments
We thank J. Drocco for useful discussions.
This work was carried out under the auspices of the 
NNSA of the 
U.S. DoE
at 
LANL
under Contract No.
DE-AC52-06NA25396.

\end{document}